\documentstyle[12pt,epsf]{article}
\renewcommand{\vec}[1]{{\bf #1}}       %%%  vectors in bold
\def\beq{\begin{eqnarray}}    %%%  begequation/eqnarray
\def\eeq{\end{eqnarray}}      %%%  endequation/eqnarray
\def\Tr{\,\mbox{Tr}\,}                  %%% Trace
\def\al{\alpha}
\def\be{\beta}

\def\ga{\gamma}
\def\de{\delta}
\def\ep{\epsilon}

\def\ze{\zeta}

\def\la{\lambda}
\def\na{\nabla}
\def\pa{\partial}

\def\si{\sigma}

\def\ph{\varphi}

\def\Ga{\Gamma}
\def\De{\Delta}

\begin{document}
%\hfill Hep-th/

\begin{center}

\vskip 2mm
%%%%%%%%%%%%%%%%%%%%%%%%%%%%%%%%%%%%%%%%%%%%%%%%%%%%%%%%%%%%%%%%
{\LARGE\sl Torsion: theory and possible observables}

\vskip 5mm

{\bf Ilya L. Shapiro} 
\footnote{ E-mail address: shapiro@ibitipoca.fisica.ufjf.br}

\vskip 2mm

{\sl Departamento de Fisica, Universidade Federal de Juiz de Fora,
36036-330, MG -- Brazil }

\vskip 1mm

{\sl and Tomsk State Pedagogical University, 634041, Tomsk, Russia }

\end{center}

\vskip 4mm

\vskip 2mm
%%%%%%%%%%%%%%%%%%%%%%%%%%%%%%%%%%%%%%%%%%%%%%%%%%%%%%%%%%%%%%%%

\noindent
{\it Abstract.}\footnote{This paper is a review based on an original 
research \cite{bush} - \cite{npt}.
To appear in "Contemporary 
Fundamental Physics", Ed. Valeri Dvoeglazov 
(Nova Science Publishers).  } $\,\,$ {\small We discuss the 
theoretical basis for the search of the possible experimental 
manifestations of the torsion field at low energies. First, the 
quantum field theory in an external gravitational field with 
torsion is reviewed. The renormalizability requires the nonminimal 
interaction of torsion with spinor and scalar (Higgs) fields. 
The Pauli-like  equation contains new torsion-dependent terms which 
have a different structure as compared with the standard electromagnetic 
ones. The same concerns the nonrelativistic equations for spin-${1}/{2}$ 
particle in an external torsion and electromagnetic fields.
Second, we discuss the propagating torsion. For the Dirac spinor
coupled to the electromagnetic and torsion field there is some additional 
softly broken local symmetry associated with torsion. As a consequence of 
this symmetry, in the framework of effective field theory, the torsion 
action is fixed with accuracy to the values of the coupling constant of 
the torsion-spinor interaction, mass of the torsion and higher derivative 
terms. The introduction of the Higgs field spoils the consistency of this
scheme, and the effective quantum field theory for torsion embedded into 
the Standard Model is not possible. The phenomenological consequences of 
the torsion-fermion interaction are drown and the case of the torsion 
mass of the Planck order is discussed.}

%%%%%%%%%%%%%%%%%%%%%%%%%%%%%%%%%%%%%%%%%%%%%%%%%%%%%%%%%%%%%%%%%%%%%%

\vskip 6mm
\noindent
{\large\bf 1. Introduction}
\vskip 2mm

Gravity with torsion and especially
the interaction of torsion with a spinor field
has attracted attention for a long time \cite{hehl,aud,rumpf,garcia}.
In a last years interest in theories with torsion has 
increased because of the success of the formal development
of string theory \cite{GSW}
which is (together with its modifications and
generalizations) nowadays regarded as the main candidate for the
unique description of all quantum fields. 
String theory predicts that the set of fields
should be extended in such a way that, along with the known 
fundamental interactions, some new ones appear. In particular,
one of these is the completely antisymmetric
torsion field which is an independent quantity serving,
along with the metric, for the description of space-time.
Probably this was the reason why in 
recent years there has been an increasing interest in possible
physical effects related to torsion. Another motivation to study the 
gravity with torsion is that it appears
naturally as a compensating field for a local gauge transformations 
\cite{kibble} (see also \cite{hehl-review} for a review of this approach
and further references). Recently there were several interesting 
developments about the possible manifestations of  torsion 
(see, for example, \cite{bbs,lame,hammond,sinryd,rysh,maroto}).
Most of these works discuss the effects of classical or quantum matter
fields on an external torsion background.

If the starting point of our consideration is
string theory, torsion should be accociated with the stress-tensor
for the antisymmetric tensor which shows up in the string effective
action. In this case, since the massive parameter of expansion
$1/\alpha'$ in string theory is of the Planck order, the only case
when the propagating torsion can be seen, is the cancellation of
massive term in the string effective action. Of course, there
is no guarantee that this cancellation doesn't take place in some
unknown version of string theory. On the other hand, if we suppose that
such a cancellation doesn't take place, then the mass of torsion is
of the Planck order and the effective field theory approach does not
apply. From physical point of view this means that torsion does´t exist
as an independent field, but only as some excitation. After the phase
transitions which breaks the string into individual field, torsion
should disappear because of its huge mass, which provides extremely fast
damping with time. The only possibility is, perhaps, that the constant
component of the torsion should not be subject of such a dumping and,
after being weakened in the inflationary epoch, can exist as a very
weak constant (in our part of the Universe) $4$-pseudovector.
Thus the discussion of the possible torsion effects should be
concentrated on the study of possibilities for the propagating torsion
and, independently, for the effects of constant background torsion to
the classical and quantum dynamics of the matter particles and fields.

The study of the renormalization of
quantum field theory in an external gravitational
field with torsion \cite{bush} (see also \cite{bush1,book}) has shown
that in the gauge theories like QED, the Standard Model or GUT's,
the interaction of matter fields with torsion has special features.
With scalars torsion can interact in a nonminimal way only.
On the other hand, if one introduces the 
interactions between spin-$1/2$, spin-0 and spin-1 fields, the
renormalizability requires nonminimal interaction with both spinors
and scalars \cite{bush1}. Thus we arrive at the necessity of  introducing
nonminimal interaction between the Dirac field and torsion, and we have to
introduce some new nonminimal parameter -- a charge, which characterizes
such an interaction.

Probably the most simple way to understand the special features
of torsion is to study nonrelativistic limit of the Dirac 
equation -- that is on the generalization of the Pauli equation for the 
case  of an external electromagnetic and torsion fields.
Here we are going to review the derivation of this approximation 
and also obtain the next to the
leading order corrections in the framework of the
Foldy-Wouthuysen transformation.
We also establish some properties of the higher order corrections,
and demonstrate some global symmetry which holds for the Dirac spinor
in external electromagnetic and torsion fields.

In the second part of this review the propagating torsion is discussed.
We prove that the action of dynamical torsion necessary
contains massive vector field and to evaluate its possible observational 
consequences. The action for the torsion
pseudovector is derived.  This action depends on two free parameters 
(one of them is torsion mass).
The study the phenomenological consequences of this action, therefore,
reduces to the search of the upper bounds for these two parameters
from the modern (in our case high-energy) experiments.
As  the result of this study we obtain these bounds on the parameters 
of the torsion action using the known data from particle physics.

The paper is organized in the following way. In the next section we
introduce basic notations, and give a very brief review of gravity with
torsion. In the next sections we
 comment on the renormalization and renormalization group for the 
gauge theories in an external gravitational field with torsion.
Some additional symmetry which holds for the spinor field
coupled to torsion is established.
In section 4 the details of the Foldy--Wouthuysen transformation
are presented. The equations of motion for spinning particles are 
discussed in section 5 and section 6 is devoted to the effective quantum
field theory for the propagating torsion.

%%%%%%%%%%%%%%%%%%%%%%%%%%%%%%%%%%%%%%%%%%%%%%%%%%%%%%%%%%%555
\vskip 6mm
\noindent
{\large\bf 2. The background notions for the gravity with torsion}
\vskip 2mm
 
In the space - time with independent metric $g_{\mu\nu}$ and torsion 
$T^\alpha_{\;\beta\gamma}$ the connection 
$\tilde{\Gamma}^\alpha_{\;\beta\gamma}$ is nonsymmetric, and
$$
\tilde{\Gamma}^\alpha_{\;\beta\gamma} -
\tilde{\Gamma}^\alpha_{\;\gamma\beta} =
T^\alpha_{\;\beta\gamma}                        
\eqno(1)
$$
If one introduces the metricity condition 
$\tilde{\nabla}_\mu g_{\alpha\beta} = 0$ where the covariant derivative 
$\tilde{\nabla}_\mu$ is constructed on the base of  
$\tilde{\Gamma}^\alpha_{\;\beta\gamma}$, 
then the following solution for connection 
$\tilde{\Gamma}^\alpha_{\;\beta\gamma}$ can be easily found
$$
\tilde{\Gamma}^\alpha_{\;\beta\gamma} = {\Gamma}^\alpha_{\;\beta\gamma} +
K^\alpha_{\;\beta\gamma}                         
\eqno(2)
$$
where ${\Gamma}^\alpha_{\;\beta\gamma}$ is standard symmetric Christoffel
symbol and $K^\alpha_{\;\beta\gamma}$ is contorsion tensor
$$
K^\alpha_{\;\beta\gamma} = \frac{1}{2} \left( T^\alpha_{\;\;\beta\gamma} -
T^{\;\alpha}_{\beta\;\gamma} - T^{\;\alpha}_{\gamma\;\beta} \right)
\eqno(3)
$$
It is convenient to divide the torsion field into three irreducible components
which are: the trace $T_{\beta} = T^\alpha_{\;\beta\alpha}$, the pseudotrace
$\;S^{\nu} = \varepsilon^{\alpha\beta\mu\nu}T_{\alpha\beta\mu}\;$ 
and the tensor
$\;q^\alpha_{\;\beta\gamma}\;$, where the last satisfies two conditions
$$ 
q^\alpha_{\;\beta\alpha} = 0,\;\;\;\;\; \;\;\;\; 
\varepsilon^{\alpha\beta\mu\nu}q_{\alpha\beta\mu} =0
$$
Then the torsion field can be written in the form
$$
T_{\alpha\beta\mu} = \frac{1}{3} \left( T_{\beta}\,g_{\alpha\mu} -
T_{\mu}\,g_{\alpha\beta} \right) - 
\frac{1}{6} \varepsilon_{\alpha\beta\mu\nu}
S^{\nu} + q_{\alpha\beta\mu}                
\eqno(4)
$$
The curvature with torsion is constructed as usual 
$[\tilde{\nabla}_\mu , \tilde{\nabla}_\nu]\Phi^A    
 ={\tilde{R}^A}_{B\,\mu\nu}\Phi^B\,\,$ 
and it has obvious relation with the Riemannian
curvature 
${\tilde{R}^A}_{B\,\mu\nu} = {R^A}_{B\,\mu\nu} + 
torsion\,\,terms$.

%%%%%%%%%%%%%%%%%%%%%%%%%%%%%%%%%%%%%%%%%%%%%%%%%%%%%%%%%%%%
Consider the Dirac field $\psi$ 
in an external gravitational field with torsion.
The standard way to introduce the minimal interaction
with external fields is the substitution of the partial 
derivatives $\partial_\mu$ by the covariant ones. 

The covariant derivatives
of the spinor field $\psi$ are defined as 
$$
\tilde{\nabla}_\mu \psi = \partial_{\mu}\psi + 
\frac{i}{2}{\tilde{w}}_\mu^{\; a b}
\sigma_{a b}\psi
$$
$$
\tilde{\nabla}_\mu \bar{\psi} = \partial_{\mu}\bar{\psi} - 
\frac{i}{2}{\tilde{w}}_\mu^{\;a b}\bar{\psi}\sigma_{a b}       
\eqno(5)
$$
where $\tilde{w}_\mu^{\;a b}$ 
are the components of spinor connection. We use 
standard representation for the Dirac matrices.
$$
\beta = \gamma^0 = \left(\matrix{1 &0\cr 0 &-1\cr} \right)
\,,\,\,\,\,\,\,\,\,\,\,\,
\vec{\alpha} = \gamma^0 \vec{\gamma} = 
\left(\matrix{0 &\vec{\sigma}\cr \vec{\sigma} & 0\cr} \right)
$$
$$
\gamma_5 = \gamma^0 \gamma^1 \gamma^2 \gamma^3, \;\;\;\;
\sigma_{a \; b} = \frac{i}{2}(\gamma_a \gamma_b - \gamma_b \gamma_a)
$$
The verbein $e_\mu^a$ obeys the equations $e_\mu^a e_{\nu a} = g_{\mu\nu}\;$, 
$e_\mu^ae^{\mu b} = \eta^{ab}$ where $\eta^{ab}$ is the Minkowsky metric. The
gamma matrices in curved space - time are introduced as $\gamma^\mu =
e_a^\mu \gamma^a$ and obviously satisfy the metricity condition 
$\tilde{\nabla}_\mu \gamma^{\beta} = 0$ . 
The condition of metricity enables one to find the explicit expression 
for spinor connection which agrees with (2). 
$$
\tilde{w}_\mu^{\;a b} = \frac{1}{4} (e_\nu^b \partial_\mu e^{\nu a} -
e_\nu^a \partial_\mu e^{\nu b}) + \bar{\Gamma}^\alpha_{\;\nu\mu}
(e^{\nu a}e_\alpha^b - e^{\nu b}e_\alpha^a)         
\eqno(6) 
$$
%%%%%%%%%%%%%%%%%%%%%%%%%%%%%%%%%%%%%%%%%%%%%%%%%%%%%%%%%%%
%%%%%%%%%%%%%%%%%%%%%%%%%%%%%%%%%%%%%%%%%%%%%%%%%%%%%%%%%%%

Then the action of spinor field
minimally coupled with torsion is written the form
$$
S = \int d^4 x \;\sqrt{-g} 
\; \{ \frac{i}{2}\bar{\psi}\gamma^\mu \tilde{\nabla}
_\mu \psi - \frac{i}{2}\tilde{\nabla}_\mu\bar{\psi}\gamma^\mu\psi +
m\bar{\psi}\psi \}           
\eqno(7)
$$
where $m$ is the mass of the Dirac field and 
%%%%%%%%%%%%%%%%%%%%%%%%%%%%%%%%%%%%%%%%%%%%%%%%%%%%%%%%%%%%%%%%%
%Further we shall consider only the 
%torsion effects and therefore restrict ourselves by the only special case 
%of flat metric. So we put $g_{\mu\nu} = \eta_{\mu\nu}$ but keep 
%$T^\alpha_{\;\beta\gamma}$ arbitrary. 
%%%%%%%%%%%%%%%%%%%%%%%%%%%%%%%%%%%%%%%%%%%%%%%%%%%%%%%%%%%%%%%%%
The expression (7) can be rewritten in the form
$$
S = \int d^4 x\,\sqrt{-g}\,\{i\bar{\psi}\gamma^\mu(\nabla_\mu+\frac{i}{8}\gamma_5
S_\mu)\psi+m\bar{\psi}\psi\}                                     
\eqno(8)
$$
where $\nabla$ is Riemann covariant derivative (without torsion).

One can see that the spinor field minimally interacts only with the 
pseudovector $S_\mu$ part of the torsion tensor. The nonminimal 
interaction is more complicated. On the dimensional grounds one can 
introduce the nonminimal coupling of the form 
$$
S = \int d^4 x\,\sqrt{-g}\,\{i\bar{\psi}\gamma^\mu(\partial_\mu 
+i\eta_1\gamma_5S_\mu+i\eta_2T_\mu)\psi+m\bar{\psi}\psi\}        
\eqno(9)
$$
Here $\eta_1,\eta_2$ are the dimensionless parameters of the
nonminimal coupling of spinor fields with torsion. The minimal
interaction corresponds to the values 
$\eta_1 = \frac{1}{8},\;\;\eta_2 = 0$.

The introduction of the nonminimal interaction seems to be artificial
since on the classical level one can explain the use of a nonminimal
action only as an attempt to explore the most general case. However
the situation is different in quantum region where the nonminimal
interaction is necessary condition of consistency of the theory. 
The reason is the following. It is well-known that the interaction 
of quantum fields leads to the divergences and therefore the
renormalization is needed. The requirement of the multiplicative
renormalizability makes us to introduce the nonminimal interaction of 
torsion with spinor and scalar fields.

%%%%%%%%%%%%%%%%%%%%%%%%%%%%%%%%%%%%%%%%%%%%%%%%%%%%%%%%%%%
%%%%%%%%%%%%%%%%%%%%%%%%%%%%%%%%%%%%%%%%%%%%%%%%%%%%%%%%%%%
With the scalar field $\,\ph\,$ torsion may interact  only 
nonminimally,
because $\,{\tilde  {\na}}\ph = \pa\ph\,$. The action of free scalar
field including the nonminimal interaction with antisymmetric
torsion has the form
$$
S_{sc} = \int d^4x\,\sqrt{-g}\,
\{\frac12\,g^{\mu\nu}\,\pa_\mu\ph\,\pa_\nu\ph
+\frac12\,m^2\ph^2 + \frac12\,\xi_i P_i\ph^2 \}
\eqno(10)
$$
where $\,\xi_i\,$ are new nonminimal parameters and
$$
P_1 = R
,\,\,\,\,\,\,\,\,\,
P_2 = \na_\al T^\al
,\,\,\,\,\,\,\,\,\,
P_3 = T_\al T^\al
,\,\,\,\,\,\,\,\,\,
P_4 = S_\al S^\al
,\,\,\,\,\,\,\,\,\,
P_3 = q_{\al\be\ga} q^{\al\be\ga}\,.
\eqno(11)
$$

We accept that the gauge vector field does not interact with 
torsion at
all, because such an interaction, generally, contradicts to the gauge
invariance. This can be easily seen from the relation
$$
{\tilde \na}_\mu A_\nu - {\tilde \na}_\nu A_\mu =
{\pa}_\mu A_\nu - {\pa}_\nu A_\mu + K^\la_{\;\;\mu\nu}\,A_\la.
\eqno(12)
$$
The nonminimal interaction with abelian vector field may be indeed
implemented in the form of the surface term
$$
S_{n-m,vec}\,=
\,i\,\al\,\int d^4x\,\ep^{\al\be\ga\si}\,F_{\al\be}\,S_{\ga\si}.
\eqno(13)
$$
Other nonminimal terms are also possible for the general
torsion but they are relevant only for the nonzero
$T_\mu$ and $q_{\al\be\ga}$ components of the torsion tensor and 
thus we are not interested in them.

%%%%%%%%%%%%%%%%%%%%%%%%%%%%%%%%%%%%%%%%%
\vskip 6mm
\noindent
{\large\bf 3. Renormalization in curved 
space-time with torsion}
\vskip 2mm

The renormalization of quantum field theories in curved space-time with
torsion has been discussed in full details in the book \cite{book}, and
therefore there are no reasons to reproduce these details here. Thus we
only establish the main qualitative results which will prove important in 
what follows. 

If the quantum theory
contains scalar and spinor fields linked by the Yukawa interaction
$\,h\ph{\bar \psi}\psi\,$, then the nonminimal parameters $\,\eta_1, xi_4\,$ 
are necessary for the renormalizability, because there are diagrams which 
lead to the corresponding divergences. As a result the nonminimal 
parameters become effective charges and their running with scale
is governed by the corresponding renormalization group equations.
For the quantum field theory
in the external torsion field the renormalization of the
parameters $\,\eta_1,\xi_4\,$ is universal. In particular, the $\be$-function 
for the nonminimal parameter $\eta_1$ has the form
$$
\be_{\eta_1} = \frac{C}{(4\pi)^2}\,h^2\,\eta_1\,,
\eqno(14)
$$
 where model-dependent $C$ is always positive.
\vskip 1mm

Despite the nonminimal fermion action contains
two dimensionless nonminimal parameters
$\,\eta_1,\eta_2\,$ we shall use only one of them
and put $\,\eta_2=0\,$. 
Reasons: (i) The minimal interaction includes only $\eta_1$
term, therefore only $\,\eta_1\,$ as an essential
parameter. (ii) The
$\eta_2$-term looks very similar to the electromagnetic interactions.
and it can be revoked by simple redefinition of the 
variables and constants.
(iii) The string-induced action depends on the completely
antisymmetric torsion which is equivalent to $\,S_\mu\,$,
so one needs only $\eta_1,\xi_4$ and $\xi_1$.
Below we denote $\eta_1=\eta$ and always take $\eta_2=0$.
%%%%%%%%%%%%%%%%%%%%%%%%%%%%%%%%%%%%%%%%%%%%%%%%%

\vskip 6mm
\noindent
{\large\bf 4. Equation for spinor field in the
nonrelativistic approximation}
\vskip 2mm

Adding usual electromagnetic interaction to the spinor field action
we get:
$$
i \hbar\frac{\partial \psi}{\partial t} = \{ c\vec{\alpha}\vec{p}-
e\vec{\alpha}\vec{A} - \eta_1 \vec{\alpha}\vec{S}\gamma_5 
+ e\Phi + \eta_1 \gamma_5 S_0 + m c^2 \beta \}\psi   
\eqno(15)
$$
Here the dimensional constants $\hbar$ and $c$ are taken into account,
and we divide 
$\;\;A_\mu = (\Phi, \vec{A})\,,\;\;\; S_\mu = (S_0, \vec{S})$.
Following standard procedure one has to write 
$$
\psi = \left(\matrix{\varphi\cr\chi\cr}\right)
exp( \frac{imc^2t}{\hbar})\,,                                  
\eqno(16)
$$
that gives
$$
(i\hbar\frac{\partial}{\partial t}- 
\eta_1 \vec{\sigma}\cdot \vec{S} - e\Phi )\varphi
= (c\vec{\sigma}\cdot\vec{p} - e\vec{\sigma}\cdot \vec{A} - \eta_1 S_0 )\chi
\eqno(17a)
$$
and
$$
(i\hbar\frac{\partial}{\partial t}- 
\eta_1 \vec{\sigma}\cdot\vec{S} - e\Phi + 2mc^2 ) \chi 
= (c\vec{\sigma}\cdot\vec{p} - e\vec{\sigma}\cdot\vec{A} - \eta_1 S_0 )\varphi
\eqno(17b)\,.
$$
Within the nonrelativistic approximation $\chi \ll \varphi$. 
Now we keep only the term $2mc^2\chi$ in the left side of (17a) and
then it is possible to find $\chi$ from (17b). 
In the leading order in
$\frac{1}{c}$ we meet the following equation for $\varphi$.
$$
i \hbar\frac{\partial \varphi}{\partial t} = 
\{ \eta_1 \vec{\sigma}\cdot\vec{S} + e\Phi 
+ \frac{1}{2mc^2\chi} (c\vec{\sigma}\cdot\vec{p} - e\vec{\sigma}\cdot\vec{A} - 
\eta_1 S_0 )^2 \} \varphi\,.
\eqno(18)
$$
The last can be easily rewritten in the Schroedinger form
$$
i\hbar\frac{\partial \varphi}{\partial t} = 
\hat{H} \varphi\,,                                                   
\eqno(19) 
$$
where the Hamiltonian has the form \cite{bbs}
$$
\hat{H} = \frac{1}{2m} \vec{\pi}^2 + B_0 + \vec{\sigma}\cdot\vec{Q} \,,
$$
$$
\vec{\pi} = \vec{P} - \frac{e}{c}\vec{A} - 
\frac{\eta_1}{c}\vec{\sigma}S_0\,,
$$
$$
B_0 = e \Phi - \frac{1}{mc^2}\eta_1^2 S_0^2\,,
$$
$$
\vec{Q} = \eta_1\vec{S} + \frac{\hbar\,e}{2mc}\,\vec{H} \,.
\eqno(20)
$$
Here $\vec{H} = rot\vec{A}$ is the magnetic field strength.
This equation is the analog of the Pauli equation in the
more general case of an external torsion and electromagnetic fields. 
Thus, there is a big difference
between the torsion and the electromagnetic terms. For instance, the term 
$- \frac{1}{mc}\eta_1 S_0 \vec{p}\cdot\vec{\sigma}\;$ does not have the
analogies in quantum electrodynamics.

%%%%%%%%%%%%%%%%%%%%%%%%%%%%%%%%%%%%%%%%%%%%%%%%%%%%%%%%%%%%%%%%%%%%%%%%%%%%%

A little bit more complicated approach comes from the 
{\sl Foldy-Wouthuysen transformation} \cite{rysh}.
The initial Hamiltonian has the form:
$$
H = \be m + {\cal E} + {\cal G}
\eqno(21)
$$
where
$\,\,
 {\cal E} = e\,A_0 - \eta\,\ga_5\,{\vec {\al}}\,{\vec S}
\,,\,\,\,\,\,{\cal G} = {\vec {\al}}
\, \left({\vec p} - e {\vec A} \right)
+ \eta\,\ga_5\,S_0\,\,\,\,$
are the even and odd parts of the Hamiltonian,
and we have used the units $c={\hbar} = 1 $.

Our purpose is to find a unitary transformation which separates
"small" and "large" components of the Dirac spinor.
One can easily see that ${\cal E}$ and ${\cal G}$ given above
obey the relations $\,\,{\cal E}\,\be = \be\, {\cal E}
\,,\,\,{\cal G} \,\be = - \be\, {\cal G}\,$
and therefore one can safely use the standard prescription for
the lowest-order approximation for ${\cal S}$:
$\,{\cal S} = - \frac{i}{2m}\,\be \, {\cal G}\,$. Thus
$$
H' = e^{i{\cal S}}
\,\left( H - i \,\partial_t\right)\,e^{-i{\cal S}}
\eqno()
$$
where ${\cal S}$ has to be chosen in an appropriate way. To find
${\cal S}$ and $H'$ in a form of the nonrelativistic expansion,
one has to take ${\cal S}$ of order $1/m$.
Then the standard result is
$$
H' = H + i \,\left[{\cal S},H \right] -
\frac12\,\left[{\cal S}, \left[{\cal S},H \right] \right] -
\frac{i}{6}\,\left[{\cal S}, \left[{\cal S}, \left[{\cal S},H
\right]  \right] \right] 
$$
$$
+\frac{1}{24}\,\left[{\cal S}, \left[{\cal S}, \left[{\cal S},
\left[{\cal S},H \right] \right]  \right] \right] 
- {\dot {\cal S}}
- \frac{i}{2}\,\left[{\cal S}, {\dot {\cal S}} \right] +
\frac{1}{6}\,\left[{\cal S}, \left[{\cal S}, {\dot {\cal S}} \right]\right]
+ ...
\eqno(22)
$$
Now we take
$$
H' = \be m + {\cal E}' + {\cal G}'
\eqno(23)
$$
where ${\cal G}'$ is of order $1/m$, and one has to perform
second FW transform with
${\cal S}' = - \frac{i}{2m}\,\be {\cal G}'$. This leads to the
$$
H'' = \be m + {\cal E}' + {\cal G}''
\eqno(24)
$$
with ${\cal G}'' \approx 1/m^2$.
%%%%%%%%%%%%%%%%%%%%%%%%%%%%%%%%%%%%%%%%%%%%%%%%%%%%%%%%%%%%%%%%%%

The third FW with
${\cal S}'' = - \frac{i}{2m}\,\be {\cal G}''$ removes odd operators
in the given order of the nonrelativistic
expansion, so that we finally obtain the usual result
$$
H''' = \be ( m + \frac{1}{2m}{\cal G}^2 -
\frac{1}{8m^3}\,{\cal G}^4 ) + {\cal E}
- \frac{1}{8m^2}\left[{\cal G},\, (
\left[{\cal G}, {\cal E}\right] + i{\dot {\cal G}})\right]
$$
Substituting here ${\cal E}$ and ${\cal G}$
after some algebra we arrive at the final form of the Hamiltonian
$$
H''' = \be \left[ m + \frac{1}{2m}
\left( {\vec p}
- e {\vec A} + \eta S_0 {\vec {\si}} \right)^2 -
\frac{1}{8m^3}\,{\vec p}^4 \right] + eA_0 -
$$$$
- \eta \left( {\vec {\si}}\cdot{\vec S} \right)
- \frac{e}{2m} {\vec {\si}} \cdot {\vec B} - \frac{\eta^2}{m}\,\be S_0^2 -
$$$$
- \frac{e}{8m^2}\,\left[ \na {\vec E} +
i{\vec {\si}}\cdot \left(\na\times {\vec E}\right)
+ 2 {\vec {\si}}\cdot\,\left({\vec E} \times {\vec p}\right) \right]+
$$$$
+\frac{\eta}{8m^2}\,\left[
- {\vec {\si}}\cdot\nabla {\dot S}_0 + \{p_i,\{p^i,
\left({\vec {\si}}\cdot{\vec S}\right)\}\} -
\right.
$$
$$
\left.
- 2\, \left(\na\times{\vec S}\right)\cdot {\vec p} +
2i \left({\vec {\si}}\cdot\nabla\right)
\left({\vec S}\cdot  {\vec p}\right)
+ 2i\,\left({\na}{\vec S}\right)\,\left({\vec {\si}}\cdot{\vec p}\right)
\right\}
\eqno(25)
$$
Here the approximation is as usual
for the electromagnetic case; that is we keep terms 

$\,\,(kinetic \,\, energy)^3\,$ and
$\,(kinetic \,\, energy)^2\cdot(potential \,\, energy)$.

One can indeed proceed in this way and get separated Hamiltonian
with any given accuracy in $\,1/m$.
We remark that the first five terms of (25) 
reproduce the Pauli-like equation with torsion.

%%%%%%%%%%%%%%%%%%%%%%%%%%%%%%%%%%%%%%%%%%%%%%%%%%%%%%%%%%%%%%

\vskip 6mm
\noindent
{\large\bf 5. The motion of spin-${1}/{2}$ particle on 
an external torsion and electromagnetic background}
\vskip 2mm

Let us start from the Pauli-like equation.
$$
H = \frac{1}{2m} \vec{\pi}^2 + B_0 + \vec{\sigma}\cdot\vec{Q} 
\eqno(26)
$$
where $\vec{\pi}, B_0, \vec{Q}$ are defined in (16) and 
$\vec{\pi} = m \vec{v}$ 
and $\vec{v} = \dot{\vec{x}}$  
is the velocity of the particle. 
The expression for the
 canonical conjugated momenta $\vec{p}$ is
$$
\vec{p} = m\vec{v} + \frac{e}{c}\vec{A} + 
\frac{\eta_1}{c}\vec{\sigma}S_0                             
$$
and  $\vec{\sigma}$ is the coordinate corresponding to spin.

Let us now 
introduce the operators of $\hat{x}_i$, $\hat{p}_i$,
spin $\hat{\sigma}_i$ and 
input the equal - time commutation relations:
$$
\left[\hat{x}_i, \hat{p}_j\right] = i\hbar \delta_{ij}, \;\;\;\;\;
\left[\hat{x}_i, \hat{\sigma}_j \right] = 
\left[\hat{p}_i, \hat{\sigma}_j\right] = 0, \;\;\;\;\;
%$$
%$$
\left[\hat{\sigma}_i,\hat{\sigma}_j \right] = 
2i\varepsilon_{ijk} \hat{\sigma}_k
$$
The Hamiltonian operator $\bar{H}$ 
is easily
constructed in terms of the operators $\hat{x}_i, \hat{p}_i, \hat{\sigma}_i$ 
and then these operators yield the equations of motion
$$
i\hbar \frac{d\hat{x}_i}{dt} = \left[\hat{x}_i, H \right],
\,\,\,\,\,\,\,
i\hbar \frac{d\hat{p}_i}{dt} = \left[\hat{p}_i, H \right],
\,\,\,\,\,\,\,
i\hbar \frac{d\hat{\sigma}_i}{dt} = \left[\hat{\sigma}_i, H \right]\,.
\eqno(27)
$$
%%%%%%%%%%%%%%%%%%%%%%%%%%%%%%%%%%%%%%
After deriving the commutators in (27) we arrive at the 
explicit form of the operator equations of motion. Now we can omit
all the terms which vanish when $\hbar \rightarrow \; 0$. 
$$
\frac{d\vec{x}}{dt} = \frac{1}{m} \left( \vec{p} - \frac{e}{c}\vec{A} - 
\frac{\eta_1}{c}\vec{\sigma}S_0 \right) = \vec{v},
\eqno(28a)
$$
\vskip 1mm
$$
\frac{d\vec{v}}{dt} = e\vec{E} + \frac{e}{c}\left[ \vec{v}\times\vec{H} \right] 
- \eta_1\left(\vec{\sigma}\cdot\nabla \right)\vec{S} - 
\eta_1\left[ \vec{\sigma}\times \na\times\vec{S} \right] 
- \frac{\eta_1}{c}\vec{\sigma}\,\frac{\partial S_0}{\partial t} 
- \frac{\eta_1}{c}S_0\frac{d\vec{\sigma}}{dt} +
$$
$$  
+ \frac{\eta_1}{c} \{ \left(\vec{v}\cdot\sigma\right) \na S_0 - 
\left(\vec{v}\cdot \na S_0 \right)\vec{\sigma} \} 
+\frac{1}{mc^2}\;\eta_1^2\; \na (S_0^2)\,,                   
\eqno(28b)
$$
$$
\frac{d\vec{\sigma}}{dt} = \left[ \vec{R}\times\vec{\sigma} \right]
$$
\vskip 1mm
$$
\vec{R} = \frac{2\eta_1}{\hbar}\left[ \vec{S} - \frac{1}{c}\vec{v}S_0 \right]
+ \frac{e}{mc}\vec{H} 
\eqno(28c)
$$

These are the
(quasi)classical  equations of motion for the particle in
an external torsion and electromagnetic fields. The operator
arrangement problem is irrelevant because of the use of 
$\hbar \rightarrow \; 0$ limit. 

The equations (28)
contain some terms which have a qualitatively new structure. 
Thus we see that the standard claim concerning magnetic field 
analogy of torsion 
effects is not completely correct, and there exist serious 
difference between magnetic field and torsion.

%%%%%%%%%%%%%%%%%%%%%%%%%%%%%%%%%%%%%%%%%%%%%%%%%%%%%%%%%%%%%%%%%
Consider some solutions of the equations of motion of a spinning particle 
in a space with torsion but without electromagnetic field.
For the cosmological reasons we are mainly
interested in the cases of constant axial vector
$S_\mu = (S_0, {\vec {S}})$. In this case the equations have the form:
$$
\frac{d{\vec {v}}}{dt}
= - \eta \,{\vec {S}}\,({\vec {v}}\cdot {\vec {\si}})
- \frac{\eta S_0}{c}\,\frac{d{\vec {\si}}}{dt}\,,
$$
$$
\frac{d{\vec {\si}}}{dt} =
+ \frac{2\eta}{\hbar}\,\left[{\vec {S}}\times {\vec {\si}}\right]
- \frac{2\eta S_0}{\hbar c}\,\left[{\vec {v}}\times {\vec {\si}}\right]\,.
\eqno(29)
$$
Consider first the case 
$S_0 = 0$ so that only ${\vec {S}}$ is present. Since ${\vec S} = const$,
we can safely put $S_{1,2}=0$. The solution for spin
can be easily found to be
${\si}_3 = {\vec {\si}}_{30} = const$ and 
$$
\si_1 = \rho\,\cos \left( \frac{2\eta S_3t}{\hbar}\right)
,\,\,\,\,\,\,\,\,\,\,\,\,
\si_2 = \rho\,\sin \left( \frac{2\eta S_3t}{\hbar}\right)\,
\eqno(30)
$$
where $\rho = \sqrt{\si_{10}^2 + \si_{20}^2}$.
For the velocity we have
$\,\,v_1=v_{10}=const,\,\,\,v_2=v_{20}=const,\,\,$
In case $\si_3 = 0$
one finds oscillating solution and
for $\si_3 \neq 0$ the solution is
$$
v_3(t) = \left[\, v_{30}\, + \,
\frac{\rho\hbar\,\left(\si_3 v_{10}\hbar -
2mv_{20}\right)}{4m^2 + \hbar^2\si_3^2}\,\right]
\;e^{-\frac{\eta S_3 \si_3}{m}\,t}\, -
$$
$$
-\,\frac{\rho\hbar}{4m^2 + \hbar^2\si_3^2}\;\left[
A_1\cos \left( \frac{2\eta S_3t}{\hbar}\right)
\,+\,A_2\sin \left( \frac{2\eta S_3t}{\hbar}\right) \,\right]
\eqno(31)
$$
with $A_1 = \si_3 v_{10}\hbar - 2mv_{20}$ and 
$A_2 = \si_3 v_{20}\hbar + 2mv_{10}$.

Thus one meets 

\noindent
(i) precession of the spin around the
direction of ${\vec S}$  

\noindent
(ii) oscillation of the particle velocity
in this same direction is accompanied (for $\si_3\neq 0$ )
by the exponential damping of the initial velocity in
this direction.

%%%%%%%%%%%%%%%%%%%%%%%%%%%%%%%%%%%%%%%%%%%%%%%%%%%%%%%%%%%%%%%%%%
Consider
another special case ${\vec {S}} = 0$, which is the form
of the torsion field motivated by isotropic cosmological models
(see, for example, \cite{book} and references there).
Then the equations of motion have a form 
$$
\frac{d{\vec {v}}}{dt}\,
= \,- \frac{\eta S_0}{c}\,\frac{d{\vec {\si}}}{dt} \,= \,
\frac{2\eta^2 S_0^2}{c \hbar}\,
\left[{\vec {v}}\times {\vec {\si}}\right]
\eqno(32)
$$
Despite the fact that those
 equations formally look like a nonlinear system
of equations for six unknowns, they can be integrated immediately
if we notice that the time variations of the variables do not affect
the vector product. Hence the general solutions are
$$
{\vec {v}}(t) = {\vec {v}}_0 +
\left[{\vec {v}}_0\times {\vec {\si}}_0\right]\,
\frac{2\eta^2S_0^2}{c\hbar}\,t\,,
$$
$$
{\vec {\si}}(t) = {\vec {\si}}_0 -
\left[{\vec {v}}_0\times {\vec {\si}}_0\right]\,
\frac{2\eta S_0}{\hbar}\,t\,.
\eqno(33)
$$ 
Thus the motion of such a particle is
a motion with constant acceleration. This is possible, in the presence
of torsion, for electrically neutral particles with spin.

Some remark is in order.
The torsion field is supposed to act on the spin of particles
but not on their angular momentum. Therefore a motion
like the one described above will occur for individual electrons or 
other particles with spin as well as for macroscopic bodies with
a nonzero overall spin orientation. However it does not occur for
the (charged or neutral) bodies with a random orientation of spins.
%%%%%%%%%%%%%%%%%%%%%%%%%%%%%%%%%%%%%%%%%%%%%%%%%%%%%%%%%%%%%%
%%%%%%%%%%%%%%%%%%%%%%%%%%%%%%%%%%%%%%%%%%%%%%%%%%%%%%%%%%%%%%

\vskip 6mm
\noindent
{\large\bf 6. Effective quantum field theory and propagating torsion}
\vskip 2mm

Independent on the development of the classical and quantum
field theory in an external torsion field it is important to
 establish the form of the action for the torsion itself and to
study the possible experimental effects of dynamical torsion.
There can be very different approaches to the construction of the
torsion action (see, for example, \cite{kibble,seni}). 
We shall consider the construction of the effective quantum field theory 
\cite{weinberg} for dynamical (propagating) torsion, and establish the 
torsion action using the consistency of this theory as a criterion. 
In this case the consistency conditions include unitarity of the S-matrix 
and the gauge-invariant renormalizability, but not the power-counting
renormalizability. The procedure of formulating of effective quantum 
field theory for the new type of interaction looks as follows:
\vskip 1mm

i) One has
to establish the field content of the dynamical torsion theory and
the form of interactions between torsion and other fields. 
\vskip 1mm

ii) It is necessary to take into account the symmetries of this 
interactions
and formulate the action in such a way that the resulting theory
is unitary and renormalizable as an effective field theory.
\vskip 1mm

Indeed there is no guarantee that these requirements are consistent
with each other, but the inconsistency may only indicate that some
symmetries are lost or that the consistent theory with the given
particle content is impossible.
For simplicity we consider below only flat metric.
%%%%%%%%%%%%%%%%%%%%%%%%%%%%%%%%%%%%%%%%%%%%%%%%%%%%%%%%%%%%%%%%%%%%%

Let us start from the
action of the fermion coupled to vector and torsion 
$$
S_{1/2}= i\,\int d^4x\,{\bar \psi}\, \left[
\,\ga^\al \,\left( {\pa}_\al - ieA_\al + i\,\eta\,\ga_5\,S_\al\,\right)
- im \,\right]\,\psi
\eqno(34)
$$
In case of the action (34) there are two gauge symmetries, 
and second of them is softly broken.
$$
\psi' = \psi\,e^{\al(x)}
,\,\,\,\,\,\,\,\,
{\bar {\psi}}' = {\bar {\psi}}\,e^{- \al(x)}
,\,\,\,\,\,\,\,\,
A_\mu ' = A_\mu - {e}^{-1}\, \pa_\mu\al(x)
$$
$$
 \psi' = \psi\,e^{\ga_5\be(x)}
,\,\,\,\,\,\,
{\bar {\psi}}' = {\bar {\psi}}\,e^{\ga_5\be(x)}
,\,\,\,\,\,\,
S_\mu ' = S_\mu - {\eta}^{-1}\, \pa_\mu\be(x)
$$
This symmetry structure enables one to derive the unique form of the
torsion action.
The higher derivative terms of the action are not seen at low energies.
Thus one arrives at the expression:
$$
S_{tor} = \int d^4\,\left\{\, -a
\,S_{\mu\nu}S^{\mu\nu} + b\,(\pa_\mu S^\mu)^2
+ M_{ts}^2\,S_\mu S^\mu\,\right\}\,,
\eqno(37)
$$
where $\,S_{\mu\nu} = \pa_\mu S_\nu - \pa_\mu S_\nu\,$ and $a,b$
are some positive parameters. (37) contains both
transversal vector mode and the scalar mode\footnote{This case has been 
considered, from different points of view, in \cite{novello,cafi}.}. 
In the $a=0$ case only the scalar mode,
and for $b=0$ only the vector mode propagate.

In the unitary
theory of the vector field the longitudinal and transversal
modes can not propagate simultaneously \cite{vector}, and therefore
one has to choose one of the parameters $a,b\,$ to be zero.
In fact the only correct choice is $b=0$, because 
the symmetry (36), which is spoiled by the massive terms only, is
preserved in the renormalization of the massless sector.
This structure of renormalization is
essentially the same as for the Yang-Mills theories with spontaneous
symmetry breaking.
On the dimensional grounds the gauge invariant counterterm 
$\,\,\int S_{\mu\nu}^2\,\,$ has to appear if we take the loop
corrections into account. We remark that the 
$\,\,\int S_{\mu\nu}^2$-term
must be included into the action even if the tree-level effects are
evaluated, if only such consideration is regarded as an approximation to
any reasonable quantum theory. Thus the torsion action is given by 
$$
S_{tor} = \int d^4 x\,\left\{\, -\frac14\,S_{\mu\nu}S^{\mu\nu}
+ M_{ts}^2\, S_\mu S^\mu\,\right\}\,.
\eqno(38)
$$
In the last expression we put the conventional coefficient $\,-1/4$
in front of the kinetic term. 

In order to illustrate that the kinetic counterterm
with $b=0$ and the massive counterterm really
show up, let us perform a simple derivation of the divergences coming
from the fermion loops.
The divergent part of the one-loop effective action is given by the
expression
$$
\Ga_{div} [A, S] = - \Tr\ln {\hat H}\mid_{div}\,,
$$ 
 where
$$
{\hat H} =
i\ga^\al \,\left({\cal D}_{\al} - im \right)
\,\,\,\,\,\,{\rm and}\,\,\,\,\,\,
{\cal D}_\al =
\pa_\al - ieA_\al + i\eta\ga_5\,S_\al\,.
\eqno(39)
$$
In order to calculate this functional determinant one can make the transformation 
which preserves covariance with respect to the derivative ${\cal D}_\al$ \cite{cogzer}.
$$
\Tr\ln {\hat H}
= \frac12\,\Tr\ln i\ga^\al \,\left({\cal D}_{\al} - im \right)
\,\cdot\,i\ga^\be \,\left({\cal D}_{\be} + im \right)=
 \Tr\ln \left(-{\hat 1}{\Box} + {\hat \Pi} \right)\,,
$$
where ${\hat \Pi} = \si^{\mu\nu}F_{\mu\nu}$.
%%%%%%%%%%%%%%%%%%%%%%%%%%%%%%%%%%%%%%%%%%%%%%%%%%%%%%%%%%%
Calculating,
using the standard Schwinger-DeWitt
technique, we arrive at the counterterms
$$
\De S [A_\mu, S_\al] 
= \frac{\mu^{D-4}}{\varepsilon} \int d^D x
\,\left\{ \frac{2e^2}{3}F_{\mu\nu}F^{\mu\nu} +
\frac{2\eta^2}{3}S_{\mu\nu}S^{\mu\nu}
+ 8m^2\eta^2S^\mu S_\mu  \right\}
$$
where $\,\varepsilon = (4\pi)^2\,(D-4)\,$ is parameter of
the dimensional regularization.
The form of the
counterterms is in perfect agreement with the
above consideration based on the symmetry transformation (36).
Namely, the one-loop divergences contain $S_{\mu\nu}^2$ and 
the massive
term while the $\,\left(\pa_\nu S^\nu\right)^2$ term is absent.
One has to notice that the topological counterterm
$$
\,\,\,ie\eta\,\int d^4x\,\ep^{\al\be\mu\nu}S_{\mu\nu}F_{\al\be}\,\,\,
$$
cancels within the covariant scheme.
This terms appears
with a nonzero coefficient 
within the noncovariant schemes. This means, in fact, the
cancellation of the one-loop contribution to the axial (or gauge)
anomaly. According to the Adler-Bardeed theorem 
in this case the anomaly is absent at higher loops too. 
The appearance of anomaly 
at one loop can, in principle, lead to the longitudinal 
counterterms in higher loops and therefore it is dangerous for the 
consistency of our model.  
Indeed for the non-abelian vector fields $\,A_\mu^a\,$ which are 
only present in SM, the
anomaly is impossible due to algebraic reasons.

%%%%%%%%%%%%%%%%%%%%%%%%%%%%%%%%%%%%%%%%%%%%%%%%%%%%%%%%%%%%%%%%%%%%%%%%
It turns out that introducing 
{\sl scalar field} $\phi$ into the fermion-torsion system is
difficult if not impossible. 
As we already know from the study of QFT on an external 
torsion background, in presence of Yukawa interactions one has to
introduce the "nonminimal" $S_\mu^2\phi^2$-vertex. Also in the torsion 
action appears additional interaction term, so it becomes
$$
S_{tor} = \int d^4x\,\{ - \frac14 S_{\mu\nu}^2
+ M_{ts}^2 S_\mu^2 - \frac{1}{24}\ze (S \cdot S)^4\} + ...
\eqno(40)
$$
Here $\,\ze\,$ is new arbitrary parameter, and coefficient
$\frac{1}{24}$ stands for the sake of convenience only. 
These two vertices can lead to problem with consistency of all 
the approach.

\begin{figure}[htb]
   \begin{center}
    \vskip -3cm\hspace*{-3cm}
    \epsfxsize=12cm\epsffile{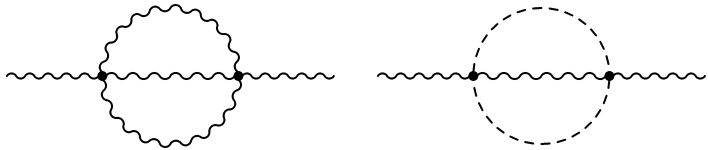}
    \vskip -12cm
\caption{\it The
wavy line is torsion propagator and dashed line -- scalar propagator}
\label{fig:dang}
\end{center}
\end{figure}

The root of the problem is that the Yukawa interaction term
$\,h\ph{\bar {\psi}}\psi$ is not invariant under the transformation
(36). Unlike the spinor mass the Yukawa constant $h$ is
massless, and therefore this noninvariance may affect the
renormalization in the massless sector of the theory. In particular,
the noninvariance of the Yukawa interaction causes the necessity
of the nonminimal scalar-torsion interaction and the self-interaction 
term in (40). At one loop there are no divergent kinetic diagrams 
with these new vertices. But at two-loop level two such diagrams 
show up, they are shown at Fig.~\ref{fig:dang}.
%%%%%%%%%%%%%%%%%%%%%%%%%%%%
%*****************************
%*****************************
Those diagrams are divergent and they can lead
to the appearance of the $\,\left(\pa_\mu S^\mu\right)^2\,$-type
counterterm. No any symmetry is seen which forbids these divergences.
Direct calculation confirms that even the $1/\ep^2$-pole
doesn't cancel.
%%%%%%%%%%%%%%%%%%%%%%%%%%%%%%%%%%%%%%%%%%%%%%%%%%%%%%%%%%%%%%%%%%%%%%%%

Let us consider two diagrams in more details.
Using the
actions of the scalar field coupled to torsion and
the torsion self-interaction (40), we arrive at the following
Feynman rules:
\vskip 1mm
\noindent
i) Scalar propagator: $\,\,\,\,\,\,\,\,\,\,\,\,
G(k) = \frac{i}{k^2+M^2}\,\,{\rm where}\,\,\,M^2=2M_{ts}^2$,
\vskip 1mm

\noindent
ii) Torsion propagator:  $\,\,\,\,\,
\,\,\,\,\,\,D_\mu^{\,\nu}(k) = \frac{i}{k^2+M^2}
\,\left(\de_\mu^{\,\nu} + \frac{k_\mu k^{\nu}}{M^2} \,\right)\,\,,$
\vskip 1mm

\noindent
iii) Torsion$^2$--scalar$^2$ vertex: $\,\,\,\,\,\,\,\,\,\,\,\,
V^{\mu\nu}(k,p,q,r)\, = \,- \,2 i\xi\,\eta^{\mu\nu}\,\,,$
\vskip 1mm

\noindent
iv) Self-interaction vertex: 
$V_{\mu\nu\al\be}
(k,p,q,r)\, =
 \,\frac{i\ze}{3}\,g^{(2)}_{\mu\nu\al\be}$
\vskip 1mm

\noindent
where 
$$
g^{(2)}_{\mu\nu\al\be} = g_{\mu\nu}g_{\al\be} +
g_{\mu\be}g_{\al\nu} + g_{\mu\al}g_{\nu\be}
$$ 
and $k,p,q,r$ denote the outgoing momenta.
\vskip 1mm

It is necessary to check the violation of 
the transversality in the kinetic 2-loop counterterms.
It turns out that it is sufficient to trace the
$\frac{1}{\ep^2}$-pole, because even this leading divergence
requires the longitudinal counterterm.
The contribution to the mass-operator of torsion from the second
diagram from is given by the following integral
$$
\Pi^{(2)}_{\al\be}(q)= -
\int \frac{d^{D} k}{(2\pi)^4} \frac{d^{D} p}{(2\pi)^4}\,
\frac{ 2\, \xi^2\,\{ \,\eta_{\al\be}\, + 
\,M^{-2}\, (k-q)_\al\,(k-q)_\be\,\}}{(p^2+
M^2)[(k-q)^2+M^2][(p+k)^2 + M^2]}\,.
\eqno(41)
$$
First one has to notice that (as in any local quantum field theory) 
the counterterms needed to subtract 
the divergences of the above integrals are local expressions, 
hence the divergent part of the above integral is finite polynomial 
in the external momenta $q^\mu$. 
In order to extract 
these divergences one can expand the factor in the
integrand into the power series in $q^\mu$
and substitute this expansion into the integral \cite{npt}. 
It is easy to see that
the divergences hold in this expansion till the order $n=8$. On the
other hand, each order brings some powers of $q^\mu$. 
The divergences of the above integral may be canceled only by adding 
the counterterms which include high derivatives.
This is a consequence of the power-counting nonrenormalizability
of the theory with massive vector fields.

To achieve the renormalizability one has to include these high 
derivative terms into the action (40). 
However, since we are aiming to construct the effective (low-energy)
field theory of torsion, the effects of the higher derivative terms are
not seen and their renormalization is not interesting for us. All we 
need are the second derivative counterterms. Hence, for our purposes
the expansion can be cut at $n=2$ rather that at $n=8$
and moreover only $O(q^2)$ terms should be kept. 
%%%%%%%%%%%%%%%%%%%%%%%%%%%%%%%%%%%%%%%%%%%%%%%%%%%%%%%%%%%%%%%%%%
Then,
using symmetry considerations, one arrives at 
$$\Pi^{(2)}_{\al\be}(q) 
= -\frac{12\,\xi^2}{(4\pi)^4\,(D-4)^2}\,q^2\,\eta_{\al\be} + ...
$$
Another integral looks a bit more complicated, but its derivation
performs in a similar way. The result reads \cite{npt}
$$
\Pi^{(1)}_{\al\la}(q)=
- \frac{\ze^2}{(4\pi)^4\,(D-4)^2}\,q^2\,\eta_{\al\la} +\, ...\,.
\eqno(42)
$$

Thus we see that both diagrams really give rise to the
longitudinal kinetic counterterm and no any simple
cancellation of these divergences is seen. On the other hand
one can hope to achieve
such a cancellation on the basis of some sofisticated symmetry. 
%%%%%%%%%%%%%%%%%%%%%%%%%%%%%%%%%%%%%%%%%%%%%%%%%%%%%%%%%%%%%%%%%%%%%

Let us compare the torsion theory with the
the usual abelian gauge theory. 
In this case the symmetry is not violated by the
Yukawa coupling, and (in the abelian case) the $\,A^2\ph^2\,$
counterterm is impossible.
The same concerns also the self-interacting $A^4$ counterterm.
The gauge invariance of the theory on quantum level is controlled
by the Ward identities. In the theory of torsion field coupled
to the MSM with scalar field 
there are no reasonable gauge identities at all. 
So there is a conflict between
renormalizability and unitarity, which reminds the
one which is well known -- the problem of massive unphysical ghosts
in the high derivative gravity. 
The difference is that in our case, unlike higher
derivative gravity, the problem appears due to the non invariance with respect
to the transformation (36).

Let us now discuss how this problem may be, in principle, solved.

i) If the torsion mass is of the Planck order then
the quantum effects of torsion should be described directly in the
framework of string theory. No any effective field theory for torsion
is possible. In this case the only visible term in the torsion action
is the massive one and torsion does not propagate at
smaller energies.

ii)
There may be a hope to impose one more symmetry which is
not violated by the Yukawa coupling. It can be, for example,
supersymmetry which mixes torsion with some vector fields of the
SM and with all massive spinor fields. In this case the
$\,\left(\pa_\mu S^\mu\right)^2\,$-type counterterm may be
forbidden by this symmetry and the conflict between renormalizability
and unitarity would be resolved.

iii) Consider the modification of SM which is free
from the fundamental scalar fields. 
%%%%%%%%%%%%%%%%%%%%%%%%%%%%%%%%%%%%%%%%%%%%%%%%%%%%%%%%%%%%%%
\vskip 3mm

Let us briefly discuss the {\sl renormalization group in the
theory with torsion}. We consider the spinor-torsion system
with an additional electromagnetic field, but without the
controversial scalar. Then the renormalization group equations for
the parameters $\,e,\eta,m,M_{ts}$ 
$$
(4\pi)^2\,\frac{de}{dt} = \frac23\,e^2
\,,\,\,\,\,\,\,\,\,\,  e(0) = e_0
$$
$$
(4\pi)^2\,\frac{d\eta}{dt} = \frac23\,\eta^2
\,,\,\,\,\,\,\,\,\,\,   \eta(0) = \eta_0
$$
$$
(4\pi)^2\,\frac{dM_{ts}^2}{dt} = 8\,m^2\,\eta^2 - 2\,M_{ts}^2
\,,\,\,\,\,\,\,\,\,\,   M_{ts}(0) = M_{ts,0}\,.
\eqno(43)
$$
We remark that the last equation demonstrates the inconsistency
of the massless or very light torsion. Even if one imposes the
normalization condition $M_{ts,0}\approx 0$ at some scale $\mu$, the
first term in this equation provides a rapid change of $M_{ts}$
such that it will be essentially nonzero at other scales. Due to the
universality of the interaction with torsion
all quarks and massive leptons should contribute to this
equation. Therefore the only way to avoid an unnaturally fast
running of $M_{ts}$ is to take its value at least of the order
of the heaviest spinor field that is $t$-quark. Hence we have some
grounds to take $M_{ts} \geq 100\,$GeV. Of course there can not be any
upper bounds for $M_{ts}$ from the RG equation.
%%%%%%%%%%%%%%%%%%%%%%%%%%%%%%%%%%%%%%%%%%%%%%%%%%%%%%%%%%%%%%%%%%%%%%%%
\vskip 6mm
\noindent
\noindent{\large\bf Phenomenology of propagating torsion}.
\vskip 2mm

The spinor-torsion interactions
enter the Standard Model as interactions of fermions with 
new axial vector field $S_{\mu}$. 
Such an interaction is characterized by the
new dimensionless parameter  -- coupling constant $\eta$. Furthermore
the mass of the torsion field $M_{ts}$ is unknown, and its
value is of crucial importance for the possible experimental
manifestations of the propagating torsion and finally for the existence
of torsion at all. Below we consider
$\,\eta\,$ and $\,M_{ts}\,$ as an arbitrary parameters and try to limit 
their values from the known experiments. Indeed we use the
renormalization group as an insight concerning the mass of torsion
but include the discussion of the "light" torsion with the mass of
the order of 1 GeV for the sake of generality. 

%%%%%%%%%%%%%%%%%%%%%%%%%%%%%%%%%%%%%%%%%%%%%%%%%%%%%%%%%%%%%%
Our strategy will be to use known experiments directed to the
search of the new interactions. We regard torsion as one of those
interactions and obtain the limits for the torsion parameters from
the data which already fit with the phenomenological considerations
\cite{npt}.
Therefore we insert torsion
into the minimal SM and suppose that the other possible new physics is 
absent. 
It is common assumption when one wants to put limits
on some particular kind of a new physics.
In this way one can  put the limits on the parameters of the
torsion action using results of various experiments. We refer the 
reader to the original work \cite{npt} for the full details,
and here present just a brief review.

It is reasonable to consider two different cases: 
\vskip 1mm
\noindent
i) Torsion is much more
heavy than other particles of SM 
\vskip 1mm
\noindent
ii) Torsion has a mass comparable to that
of other particles. In the last case one meets a propagating
particle which
must be treated on an equal footing with other constituents of the SM.
\vskip 1mm
\noindent
%%%%%%%%%%%%%%%%%%%%%%%%%%%%%%%%%%%%%%%%%%%%%%%%%%%%%%%%%%%%%%%%%

Indeed the very heavy torsion leads to the effective contact
four-fermion interactions.
Consider the case of heavy torsion in some more details. Since
the massive term dominates over the covariant kinetic
part of the action, the last can be disregarded. Then the total 
action leads to the algebraic equation of motion
for $\,S_\mu$. The solution of this equation can be substituted back to
$\,S_{1/2} + S_{tor}\,$ and thus produce the contact four-fermion
interaction term
$$
{\cal L}_{int} = - \frac{\eta^2}{M_{ts}^2}\,
({\bar \psi}\ga_5\ga^\mu\psi)\,({\bar \psi}\ga_5\ga_\mu\psi)
\eqno(44)
$$

As one can see the only one quantity which appears in this 
approach is the ratio ${M_{ts}}/{\eta}\,$ and therefore for the very 
heavy torsion field the phenomenological consequences depend only 
on single parameter.

Physical observables related with torsion depend on the two 
parameters $M_{ts}$ and $\eta$.
In the course of our study we choose, for the sake of simplicity,
all the torsion couplings with fermions to be the same $\eta$.
 This enables one to put the limits in the two
dimensional $M_{ts}\,-\, \eta$ parameter space using
the present experimental data.
We also assume that non-diagonal coupling of the torsion with the  
fermions of different families
is zero in order to avoid flavor changing neutral current problem.

\begin{figure}[htb]
   \begin{center}
    \vskip -1cm\hspace*{-3cm}
    \epsfxsize=12cm\epsffile{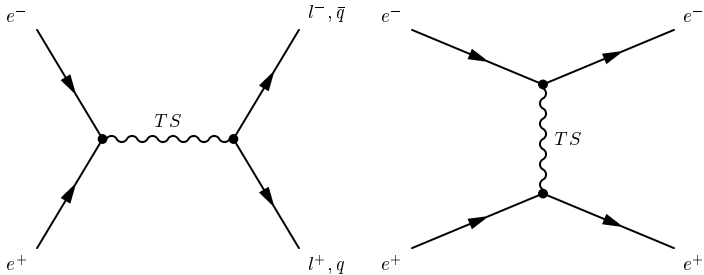}
    \vskip -12cm
\caption{\it Diagrams with torsion exchange in $e^+e^-$
collisions contributing to 
forward-backward lepton and quark asymmetry. TS indicates to the 
torsion propagator.}
    \vskip -0.5cm
\label{eediag}
\end{center}
\end{figure}

%%%%%%%%%%%%%%%%%%%%%%%%%%%%%%%%%%%%%%%%%%%%%%%%%%%%%%%%%%%%%%%%%%%
Another possibility comes from the consideration of the
the axial-vector type interactions would give rise to the 
forward-backward
asymmetry. This asymmetry has been presizely measured in the
$e^+e^- \rightarrow l^+l^- (q\bar{q})$  scattering
(here $l$ stands for tau,muon or electron) at LEP collider with the
center-mass energy equals to the Z-boson mass. The corresponding diagrams
are shown at Fig.~\ref{eediag}.  
%%%%%%%%%%%%%%%%%%%%%%%%%%%

%%%%%%%%%%%%%%%%%
Indeed any parity violating interactions give rise to the space
asymmetry and could be reviled
in a forward-backward asymmetry measurement. 
Axial-vector type interactions of torsion with matter fields is
this case of interactions.
But the  source of asymmetry also exists electroweak interactions  
because of the presence of the $\gamma_\mu\gamma_5$ structure in
the  interactions of $Z$- and $W$-bosons with fermions.
 Such an asymmetries are measured at LEP.
Presence of torsion would  change  the forward-backward asymmetry and
would brightly reviel itself since its contribution differs from the
one of other fields \cite{npt}. 
One has to notice the measured EW parameters are in a good agreement 
with the
theoretical predictions and hence the limits established on the
torsion parameters are based on the experimental errors.
%%%%%%%%%%%%%%%%%%%%%%%%%%%%%%%%%%%%%%%%%%%%%%%%%%%

The straightforward consequence of the  heavy torsion
interacting with fermion fields is 
the effective four-fermion contact interaction of leptons and quarks.
The contact interactions were widely discussed in 70th (see \cite{hehl}
for the review.)
Four-fermion interaction effectively appears
for the torsion with a mass much higher than the energy
scale available at present colliders. 
Thus it is interesting to discuss the possibility for detecting 
torsion effects in the framework of new experimental possibilities.
There are several experiments from which the constraints on the 
contact four-fermion interactions come. One can find further 
details in \cite{npt}.
The specific distinguishing feature of the contact interactions
induced by torsion is that those contact interactions are of 
axial-axial type. In \cite{npt} 
we have used the limits obtained by other authors
for this kind of interaction. 

%%%%%%%%%%%%%%%%%%%%%%%%%%%%%%%%%%%%%%%%%%%%%%%%%%%%%%%%%%%%%
The torsion with the mass
in the range of  present colliders could be produced in fermion-fermion
interactions as a resonance, decaying to fermion pair.
The most promising collider for search the signature
of such type is TEVATRON. 
	The total limits on torsion parameters 
coming from all mentioned experiments, can be found in \cite{npt}.

%%%%%%%%%%%%%%%%%%%%%%%%%%%%%%%%%%%%%%%%%%%%%%%%%%%%%%%%%%%%%
\vskip 6mm
\noindent
{\large\bf Conclusion and Acknowledgments}.
\vskip 2mm

We have reviewed the recent developments about the possible
manifestations of torsion field. In particular, the non-relativistic
approximation to the Dirac equation and the corresponding action
of particle have been derived. Also we discussed the possibility
to implement propagating torsion into the Standard Model of the
elementary particle physics, and found that this can be done, but
only for the fermion sector of the SM. The introduction of torsion
into the full SM including Higgs fields meets serious difficulties.

Author is grateful to all his colleagues, with whom he collaborated
in the study of torsion.
I am especially indebted to M. Asorey, I.L. Buchbinder,
A. Belyaev and L. Ryder for the new things I have learned from them
and with their assistance, and to  V.G. Bagrov, J.A. Helayel-Neto,
I.B. Khriplovich and T. Kinoshita for helpful discussions.

Author is also grateful to the
Physics Department at the Federal University of Juiz de Fora
for warm hospitality and to Brazilian Foundations CNPq and
FAPEMIG (Minas Gerais) for support.

%%%%%%%%%%%%%%%%%%%%%%%%%%%%%%%%%%%%%%%%%%%%%%%%%%%%%

\end{document}